\documentclass[letter]{aa}  
\usepackage{multirow} 
\usepackage{natbib}
\usepackage{amssymb,amsmath}
\usepackage{graphicx, subfigure}
\usepackage{siunitx}
\usepackage{url}
\usepackage{threeparttable}
\usepackage{txfonts}
\usepackage{ textcomp }

\usepackage{color, colortbl}
\usepackage{booktabs}
\begin{document}

   \title{Detection of highly excited OH towards AGB stars}
   \subtitle{A new probe of shocked gas in the extended atmospheres}

   \author{T. Khouri\inst{1}\thanks{{\it Send offprint requests to T. Khouri}
   \newline \email{theokhouri@gmail.com}}, L. Velilla-Prieto\inst{1}, E. De Beck \inst{1}, W. H. T. Vlemmings\inst{1},
   H. Olofsson\inst{1}, B. Lankhaar\inst{1}, J. H. Black\inst{1}, A. Baudry\inst{2}
}

\institute{Department of Space, Earth and Environment, Chalmers University of Technology, Onsala Space Observatory, 43992 Onsala, Sweden 
\and Laboratoire d'astrophysique de Bordeaux, Universit\'e de Bordeaux, CNRS, B18N, all\'ee Geoffroy Saint-Hilaire, 33615 Pessac, France 
}

  \abstract
  {}
  {We  characterise the gas in the extended atmospheres of the oxygen-rich asymptotic giant branch (AGB) stars W~Hya and R~Dor using high angular resolution
  ALMA observations.}
  { We report the detection and investigate the properties of high-excitation $\Lambda$-doubling line emission of hydroxyl (OH).}
  {The OH lines are produced very close to the central stars and seem optically thin and with no maser effect. We analyse the molecular excitation using a population diagram
  and find rotational temperatures of $\sim 2500$~K and column densities of $\sim 10^{19}$~cm$^{-2}$
  for both sources. For W~Hya, we observe emission from  vibrationally excited H$_2$O arising from the same region as the OH emission. Moreover,
  CO~${v=1, J=3-2}$ emission also shows a brightness peak in the same region.
  Considering optically thin emission
  and the rotational temperature derived for OH, we find a CO column density $\sim 15$ times higher than that of OH, within an area
  of ($92\times84$) mas$^2$ centred on the OH emission peak. These results should be considered tentative because of the simple methods employed.
  The observed OH line frequencies differ significantly from the predicted transition frequencies in the literature, and provide the
  possibility of using OH lines observed in AGB stars to improve the accuracy of the Hamiltonian used for the OH molecule.
  We predict stronger OH $\Lambda$-doubling lines at millimetre wavelengths than those we detected. These lines will be a good probe
  of shocked gas in the extended atmosphere and are possibly even suitable
  as probes of the magnetic field in the atmospheres of close-by AGB stars through the Zeeman effect.}
{}
               
\titlerunning{First detection of highly excited OH in AGB stars}
\authorrunning{T. Khouri et al.}

\maketitle

\section{Introduction}

At the ends of their lives, low- and intermediate-mass stars become very luminous and cool as they
ascend the asymptotic giant branch (AGB). They develop an extended
atmosphere where a rich molecular chemistry occurs, leading to the formation of dust and
to copious mass loss. 
Shocks happen periodically in the atmosphere because of strong stellar pulsations that take place during the AGB
phase and of the motion of large convective cells in the stellar hydrogen envelope \citep[e.g.][]{Hoefner2018}. These shocks
can dissociate molecules and have a major effect on the chemical composition of the gas \citep[e.g.][]{Cherchneff2006}. 

Hydroxyl is a very important radical for shock chemistry in the inner circumstellar envelope, and is also produced
in its outer parts through photodissociation of H$_2$O.
An alternative possibility for forming highly excited OH in the inner envelope is photodissociation of H$_2$O by ultraviolet line emission from a stellar chromosphere
or a hot binary, similar
to the mechanism discussed by \cite{Tappe2008} in the Herbig-Haro object HH 211.

The unpaired electron of OH and the nuclear magnetic dipole moment of H give rise to a rich microwave and
a near-/far-infrared line spectrum \citep[e.g.][]{Beaudet1978, Drouin2013}. Detections towards evolved stars include
ro-vibrational \citep{Hinkle1978}, spin-flip, and pure rotational \citep{Sylvester1997},
and the famous 18cm maser-type emission from the ground state of OH \citep[first reported in stars by][]{Wilson1968}.
While the ro-vibrational lines are produced close to the star by gas in the shocked atmosphere,
all the other lines observed thus far are expected to be produced in the outer envelope after the dissociation of H$_2$O.

In this letter, we report the first detection of highly excited ($E_{\rm u}/k \gtrsim 4800$~K) $\Lambda$-doubling hyperfine lines
towards three AGB stars, W~Hya, R~Dor and IK~Tau, based on observations with the Atacama Large
Millimeter/submillimeter Array (ALMA) at high angular resolution.
These lines are produced in the atmosphere and provide a new probe of gas in these inner regions.
Every spin-rotation state of OH, defined by the rotational angular momentum ($N$) and hyperfine-free angular momentum ($J$),
has two $\Lambda$ components of different parity, each of which is further
split into two hyperfine substates of total angular momentum $F=J\pm 1/2$. We detect only the main lines ($\Delta F = 0$) within different J  rotational levels.
This is expected as their Einstein-$A$ coefficients are between two and three orders of magnitude larger than those of the two satellite lines ($\Delta F = \pm 1$).
{In this letter} we give the excitation energy of lines, $E_{\rm u}/k$, in K, where $E_{\rm u}$
is the excitation energy of the upper level of the transition and $k$ the Boltzmann constant.

\section{Observations}

The first detections of OH presented in this paper were made in the observations that were part of
projects 2016.1.00004.S and 2017.1.00075.S (PI Vlemmings). The aim of these  projects was to resolve the O-rich AGB stars
W~Hya, R~Dor, $o$~Ceti, and R~Leo in ALMA Bands 4 and 6, using ALMA configurations that obtained an angular resolution $<30$~mas.
The details of the observational set-up and calibration will be presented in a forthcoming paper (Vlemmings et al., in prep.).
Generally, the observatory-provided calibration or pipeline scripts were used, after which the data were self-calibrated on the strong
stellar continuum. We first detected the two strongest OH $\Lambda$-doubling lines within the $N_J = 13_{25/2}$ state at $\sim 130.1$~GHz
around W~Hya.
Subsequently, we searched for similar OH lines at other frequencies in our data and in archival observations of close-by O-rich AGB stars
(W Hya, R Dor, $o$~Ceti, R Leo, and IK Tau). Also for these observations, typically at lower angular resolution, we applied the calibration provided
by the observatory followed by continuum self-calibration. In doing so, we discovered multiple high-quality detections of $\Lambda$-doubling line pairs
towards W Hya and R Dor, and one such line pair towards IK~Tau. The additional observations of W~Hya in which we found $\Lambda$-doubling
OH emission are from projects 2016.1.00374.S (Band 6, PI Ohnaka), and 2016.A.00029.S (Band 7, PI Vlemmings). For R~Dor, we found emission in data
from projects 2017.1.00582.S (Bands 3, 6, and 7, PI De Beck) and also, unidentified, in published data \citep{Decin2018}. The emission towards IK~Tau was
detected in the data of project 2017.1.00582.S (Band 6, PI De Beck). In the Appendix we show the spectra observed towards W~Hya and IK~Tau
in Fig.~\ref{fig:linesWHya_IKTau} and R~Dor in Fig.~\ref{fig:lines}. In Table~\ref{tab:dates}, we give the dates when each observation was obtained.
A search for OH emission in available observations of
$o$~Ceti and R~Leo was unsuccessful.

 \begin{table}[t]
\footnotesize
\setlength{\tabcolsep}{12pt}
\caption{OH lines detected towards W~Hya and R~Dor. In the second and third columns,
we indicate the transitions specifying the quantum numbers described
in the text and the positive or negative parity of the energy levels involved.
We also give the predicted line frequency ($\nu$) with corresponding uncertainty {(obtained from the JPL database, see text)},
the excitation energy of the upper level of the transition ($E_{\rm u}/k$, where $k$ is the Boltzmann constant) and
the integrated line flux ($\int F_\nu d\nu$) with corresponding uncertainty.
{We also list} the Einstein-$A$ coefficient of the transitions ($A_{\rm ul}$, {calculated from the line strengths given in the JPL database}).
The line fluxes for W~Hya are given for the large (LR) and small
(SR) regions (see text).}              
\label{tab:lines}      
\centering                                      
\begin{tabular}{@{}c @{\phantom{a}}c @{\phantom{0}}c @{\phantom{0}}c @{\phantom{0}}c @{\phantom{}} c @{\phantom{}}c@{} }
\hline
$\nu$ & $\varv, N_{J}$ & $F^{\prime}-F^{\prime\prime}$ &$\int F_\nu d\nu$ &  $E_{\rm u}/k$ & $A_{\rm ul}$\\
(MHz) & & & ($10^{-22}$ W m$^{-2}$) & (K) & ($10^{-7} $s$^{-1}$) &  \\
\hline
\\
& & & W Hya LR (SR) & & & \\
\hline
130078 (0.1)\phantom{0} & $0, 13_{\frac{25}{2}}$ & $13^- - 13^+$ & $8.0 \pm 1.5$ ($2.9 \pm 0.8$) & 4818 & 2.82 \\
130114 (0.1)\phantom{0} & $0, 13_{\frac{25}{2}}$ & $12^- - 12^+$ & $7.0 \pm 1.5$ ($3.2 \pm 0.8$) & 4818 & 2.82 \\ 
130639 (0.2)\phantom{0} & $1, 10_{\frac{21}{2}}$ & $10^+ - 10^-$  & $10 \pm 1.5$ ($3.6 \pm 0.8$) & 7930 & 8.55 \\
130655 (0.2)\phantom{0} & $1, 10_{\frac{21}{2}}$ & $11^+ - 11^-$ & $7.5 \pm 2.5$ ($2.3 \pm 0.8$) & 7930 & 8.55 \\
265734 (0.8)\phantom{0} & $0, 18_{\frac{35}{2}}$ & $18^+ - 18^-$ & $14 \pm 2 $ (--) & 8860 & 13.8 \\
265765 (0.8)\phantom{0} & $0, 18_{\frac{35}{2}}$ & $17^+ - 17^-$ & $10 \pm 3$ (--) & 8860 & 13.8 \\ 
333394 (2.0)\phantom{0} & $1, 17_{\frac{35}{2}}$ & $17^- - 17^+$ & $20 \pm 7$ ($7.0 \pm 2.0$) & 12756 & 47.2 \\ 
333412 (2.0)\phantom{0} & $1, 17_{\frac{35}{2}}$ & $18^- - 18^+$ & $24 \pm 7$ ($8.3\pm 2.0$) & 12756 & 47.2 \\ 
\\
& & & R Dor & & & \\
\hline
107037 (0.06) & $0, 12_{\frac{23}{2}}$ & $12^+ - 12^-$ & $1.85 \pm 0.17$ & 4149 & 1.79 \\
107073 (0.06) & $0, 12_{\frac{23}{2}}$ & $11^+ - 11^-$ & $1.1 \pm 0.4$ & 4149 & 1.79 \\
130078 (0.1)\phantom{0} & $0, 13_{\frac{25}{2}}$ & $13^- - 13^+$ & $< 5$ & 4818 & 2.82 \\
130114 (0.1)\phantom{0} & $0, 13_{\frac{25}{2}}$ & $12^- - 12^+$ & $< 5$ & 4818 & 2.82 \\
252127 (0.4)\phantom{0} & $0, 14_{\frac{29}{2}}$ & $14^+ - 14^-$ & $30.8 \pm 1.1$ & 5495 & 30.33 \\
252145 (0.4)\phantom{0} & $0, 14_{\frac{29}{2}}$ & $15^+ - 15^-$ & $34.5 \pm 1.5$ & 5495 & 30.34 \\ 
317397 (0.9)\phantom{0} & $0, 16_{\frac{33}{2}}$ & $16^+ - 16^-$ & $64 \pm 10$ & 7072 & 45.87 \\
317415 (0.9)\phantom{0} & $0, 16_{\frac{33}{2}}$ & $17^+ - 17^-$ & $67 \pm 10$ & 7072 & 45.87 \\
351593 (1.2)\phantom{0} & $0, 17_{\frac{35}{2}}$ & $17^- - 17^+$ & $72 \pm 10$ & 7928 & 55.0 \\
351612 (1.2)\phantom{0} & $0, 17_{\frac{35}{2}}$ & $18^- - 18^+$ & $92 \pm 20$ & 7928 & 55.0 \\
\\

\end{tabular}
\end{table}

\section{Observational results}

\subsection{W~Hya}

\begin{figure}[t]
\centering
    \includegraphics[width=9cm]{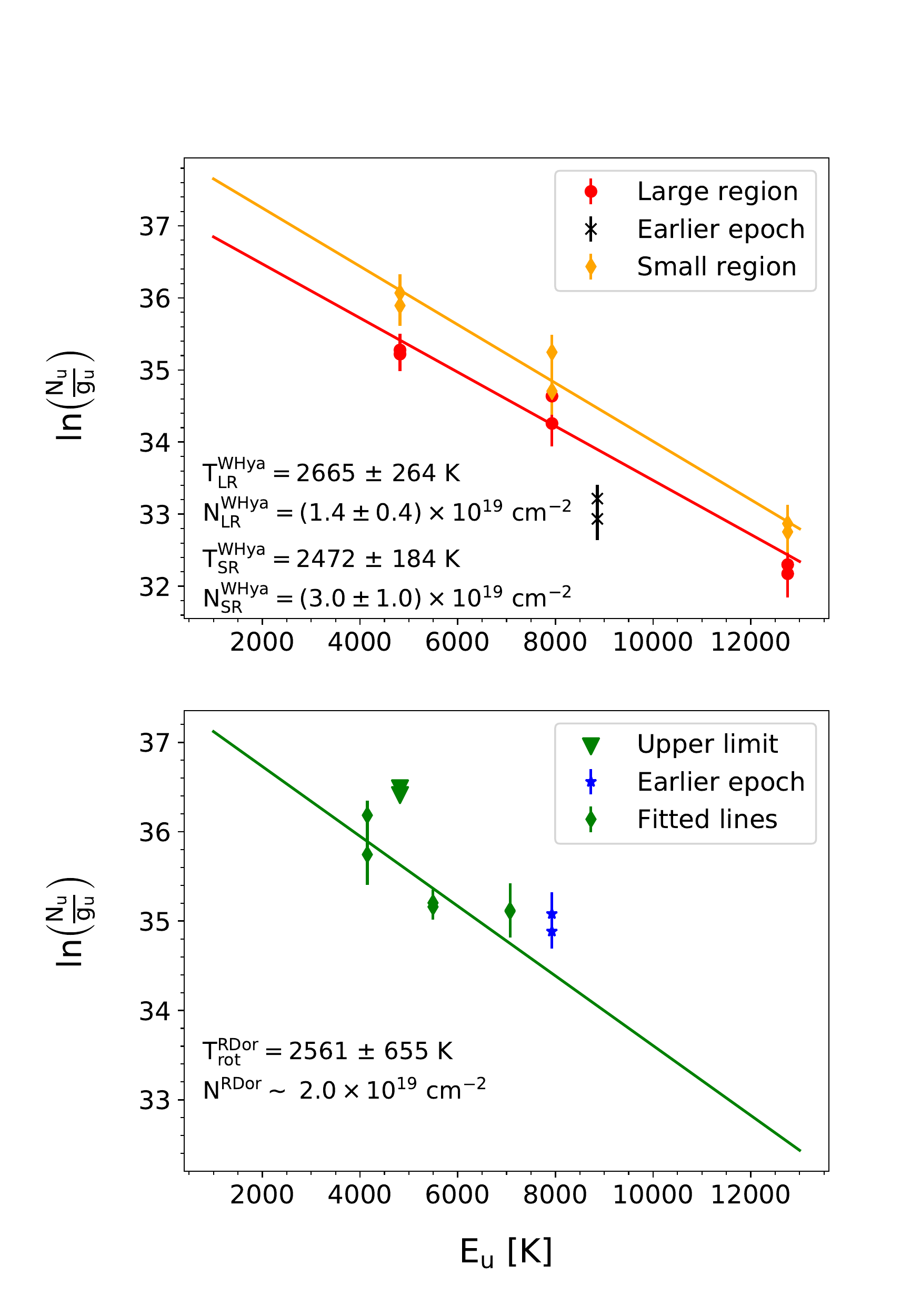}
    \caption{Population diagrams for lines detected towards W~Hya ({\it upper panel}) and R~Dor ({\it lower panel}).
    We performed the fit considering lines observed within approximately one month in time.
    For W~Hya, we performed the population diagram analysis considering line fluxes extracted from two distinct regions,
    see text for details.}
        \label{fig:popDiag}
\end{figure}

\begin{figure}[t]
\centering
    \includegraphics[width=9.5cm]{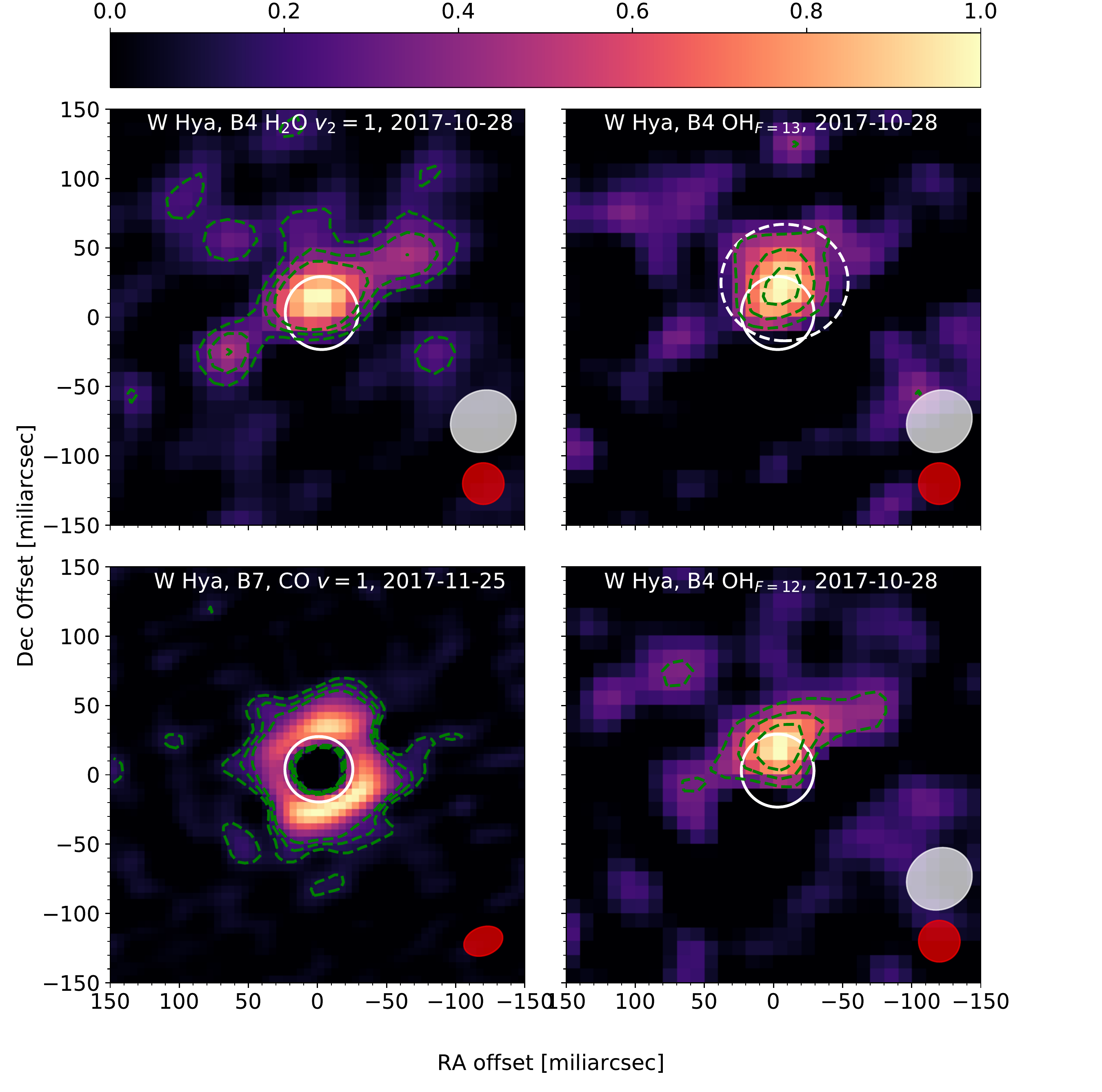}
    \caption{Emission regions observed towards W~Hya of the lines ${v_2=1, {J_{K_{\rm a},K_{\rm c}}} = 8_{8, 1} - 9_{7, 2}}$ and
${v_2=1, {J_{K_{\rm a},K_{\rm c}}} = 8_{8, 0} - 9_{7, 3}}$ of H$_2$O {(sum of the two lines)}; $\varv=0, N_J =13_{25/2}$ with $F=12-12$ and $F=13-13$ of OH;
and $\varv=1, J=3-2$ of CO.
    {The white contours show the size of the stellar disk at the appropriate frequencies (measured from fits of a uniform-disk source to the observed
    visibilities; Vlemmings et al., in preparation),} the colour map shows the
    integrated line intensity normalised to its peak value, and the dashed green contours show emission 3, 5, and 7 times
    above the root mean square noise of the given image. {The dashed white circle shows the extent of the
    small region (see text).} The grey and red ellipses show the restoring beam of the
    continuum and line images, {respectively}. For the images in band 7, the two beams are equal.}
        \label{fig:linesWHya}
\end{figure}

We detect three OH line pairs towards W Hya, in observations acquired at similar stellar pulsation phases. One pair lies in the vibrational ground state,
$\varv=0$, the other two in the first vibrationally excited state, $\varv=1$. The spread of more than 8000~K in excitation energy allows us to obtain good constraints on the column density
and rotational temperature using a population diagram \citep[Fig.~\ref{fig:popDiag}]{Goldsmith1999}.
For this analysis, we use the partition function given by \cite{Drouin2013}.
The spatially resolved observations (Fig.~\ref{fig:linesWHya}) allow us to produce population diagrams with line fluxes from two regions,
a small region (${\sim 6.1 \times 10^{-3}~{\rm arcsec}^2}$, shown in Fig.~\ref{fig:linesWHya}) centred on the emission peak,
and a larger region (${\sim 3.1 \times 10^{-2}~{\rm arcsec}^2}$) that includes the star and the circumstellar environment to the north
and that encompasses all detected emission.
The line fluxes were measured by fitting Gaussian functions to the observed spectra. The only exception are the fluxes from
lines within level $\varv, N_J = 1, 17_{35/2}$ which were obtained by integrating the spectrum numerically.
This approach was chosen because these lines are seen close to several other features and it is not easy to identify the continuum level in the immediate
vicinity of the lines.
The line fluxes obtained are given in Table~\ref{tab:lines}.
{From a population diagram analysis (Fig.~\ref{fig:popDiag}),} we find rotational temperatures for the small and large regions of $2472 \pm 184$~K and $2665 \pm 264$~K, respectively,
and corresponding OH column densities of ${(3.0 \pm 1.0) \times 10^{19}~{\rm cm}^{-2}}$
and ${(1.4 \pm 0.3) \times 10^{19}~{\rm cm}^{-2}}$.
The rotational temperatures derived are equal when considering the fit uncertainties.

We identify another $\Lambda$-doubling line of OH at $\sim 265.5$ GHz in earlier observations of W~Hya (PI Ohnaka),
at pulsation phase $\sim 0.7$. The line strength is lower than expected based on the population diagram obtained by us for the large region.
An increase in column density by a factor of $\sim 2.5$ (for the same rotational temperature) is enough
to explain the increase in line strength between epochs
0.7 and 1.0.


The lines are produced at least partially in front of the stellar disk and
in the northern region of the extended atmosphere, and thus show emission even against the stellar continuum (Fig.~\ref{fig:linesWHya}).
This suggests that at least a fraction of the OH gas has a rotational temperature higher than the continuum brightness temperature.
A weak absorption feature seems to be present in most lines and becomes clearer when we extract a spectrum
towards the star (Fig.~\ref{fig:linesWHya_IKTau}). This indicates that there is a fraction of the OH gas with rotational
temperature lower than the brightness temperature of the continuum.
These two findings can be reconciled by considering that the environment where the OH lines are produced is inhomogeneous
and that the continuum brightness temperatures at bands 4 and 7, respectively   $\sim 2500$~K and $\sim 2650$~K, 
are very similar to the rotational temperatures we derive.
If a significant fraction of
the OH molecules is seen against the stellar disk, the column densities we obtain using the population diagram analysis
will be underestimated because of absorption or weaker emission against the background of the stellar disk.

\subsubsection*{Comparison to H$_2$O and CO lines}

In the band 4 data obtained towards W~Hya we detect the ${v_2=1, {J_{K_{\rm a},K_{\rm c}}} = 8_{8, 1} - 9_{7, 2}}$ and
${v_2=1, {J_{K_{\rm a},K_{\rm c}}} = 8_{8, 0} - 9_{7, 3}}$ rotational lines of ortho- and para-H$_2$O, respectively,
which are produced in the same region as the OH lines (Fig.~\ref{fig:linesWHya}).
These two H$_2$O lines have excitation energies of $5080$~K. 
The lines have FWHM of $\sim 8$~km/s and peak flux density in the small region of 
${\sim40}$~mJy and ${\sim25}$~mJy for the
${v_2=1, {J_{K_{\rm a},K_{\rm c}}} = 8_{8, 1} - 9_{7, 2}}$ and
${v_2=1, {J_{K_{\rm a},K_{\rm c}}} = 8_{8, 0} - 9_{7, 3}}$ lines, respectively.
This implies optical depths of $\sim 0.2$, and thus optical depth effects might be important.
Given the complex excitation of H$_2$O and the small number of observed lines, we refrain from estimating the H$_2$O column density.

The extended atmosphere is traced by CO~${v=1, J=3-2}$ emission
detected at 342.648~GHz. Observations of this line acquired in December 2015 reveal an emission
region with a radius of $\sim 48$~mas centred on the star \citep{Vlemmings2017}, similar in size to that shown in the data we discuss.
\citeauthor{Vlemmings2017} modelled the emission line in these earlier observations and found the gas
mass in the emission region to be $\sim 1.4 \times 10^{-4}$~M$_\odot$.

The CO $\varv=1,J=3-2$ line shows an emission peak in the same region where the OH emission arises.
This is an indication of a larger column density of gas, higher gas temperature, or more efficient radiative excitation in this region.
To obtain an estimate of the CO column density, we extract the CO~$\varv=1, J=3-2$ line emission from the small region
used in the OH analysis. In this CO line spectrum, emission reaches a peak of 140~mJy at $\upsilon_{\rm LSR} = 39.2$~km/s,
while absorption reaches a level of $\sim -50$~mJy
at a velocity of $\sim 10$~km/s with respect to the $\upsilon_{\rm LSR}$. The CO spectrum extracted towards the star shows no
absorption at $\sim \upsilon_{\rm LSR}$ and we conclude that the line peak level is not strongly affected by absorption.
Hence, we approximate the CO line in the small region
by a Gaussian line with the observed peak value (140~mJy) and an inferred full width at half maximum (FWHM) of $\sim 9$~km/s.
We estimate the CO column density
assuming optically thin emission and considering an excitation temperature of $2000 \pm 1000$~K. {This broad range in excitation temperatures
was chosen to encompass the derived rotational temperature of OH and lower values derived from modelling earlier CO~$\varv=1, J=3-2$ observations
\citep{Vlemmings2017}}. We find
$N_{\rm CO} \sim 4^{+3}_{-1} \times 10^{20}$~cm$^{-2}$ and an optical depth at the line centre of $\sim 0.3\pm0.2$.
This result indicates a tentative CO-to-OH column density ratio of $\sim 15$ in the small region around the OH emission peak.

Considering a CO abundance relative to H$_2$ of $4 \times 10^{-4}$ {\citep[e.g.][]{Willacy1997}}, the derived CO-to-OH column density ratio suggests an OH abundance
relative to H$_2$ in the small region of $\sim 3 \times 10^{-5}$. This is in agreement with abundances calculated in post-shocked gas
under the assumption of thermal equilibrium, but a factor of 50 higher than that obtained without a thermal equilibrium assumption \citep{Cherchneff2006}.
{Assuming the depth of the small region to be comparable to its outer radius ($\sim 70$~mas),}
we derive an average H$_2$ density of $\sim 10^{10}$~cm$^{-3}$. This is too low for
collisions to dominate the excitation of CO to $\varv=1$ \citep[e.g.][]{Woitke1999}.
We note that this average density is most likely much lower than the densities where
the OH emission is produced, very close to the star.
The diluted stellar radiation field will have a temperature close to the rotational temperature we derive,  and thus  is probably the
dominant factor controlling the excitation of CO, and probably OH.
This suggests  that the OH abundance we determined  is a crude estimate and that a
proper determination must rely on non-local thermodynamic equilibrium (non-LTE) modelling of  the observed OH and CO lines.

\subsection{R~Dor}

\begin{figure*}[t]
\centering
    \includegraphics[width=15cm]{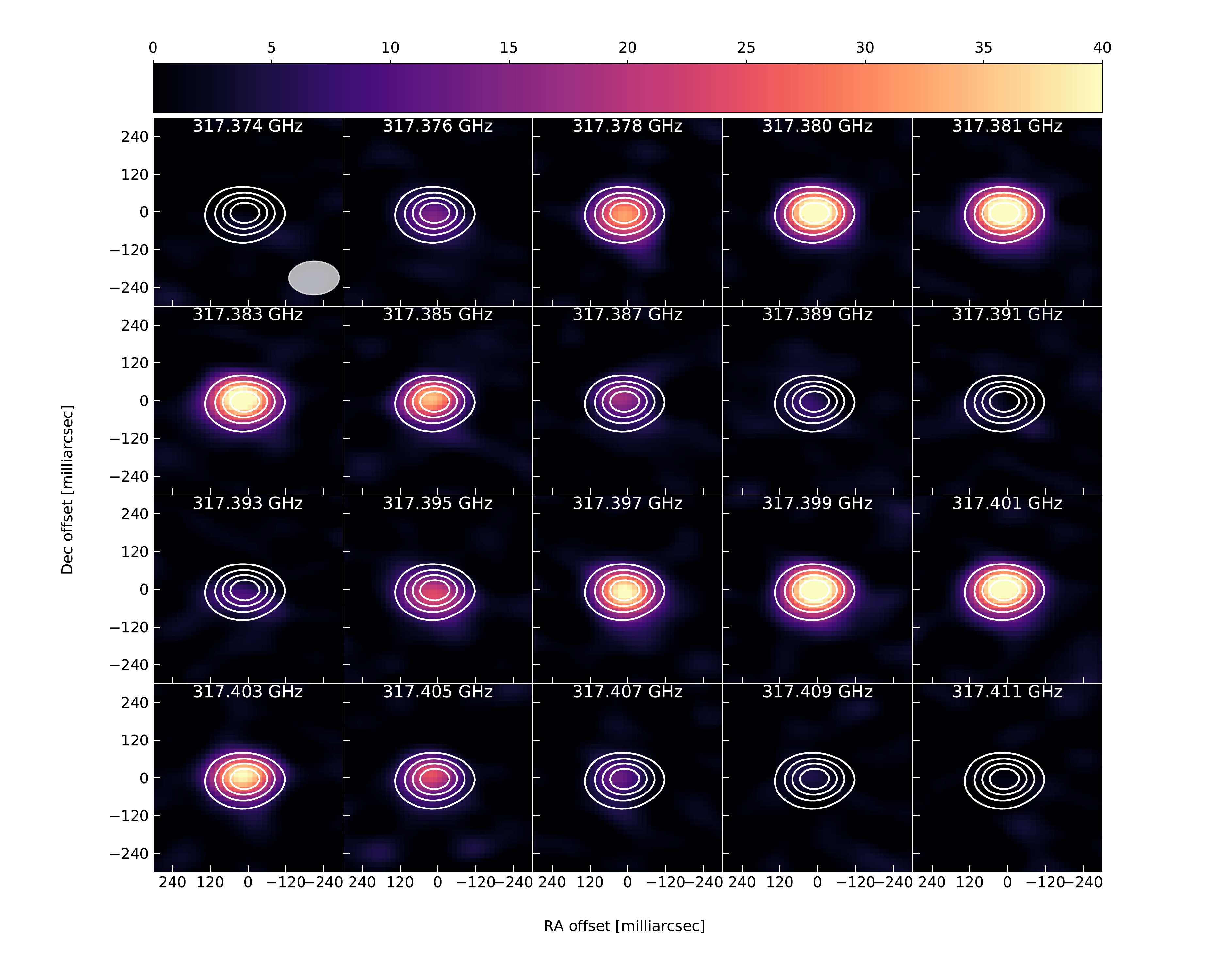}
    \caption{Channel map of the $\Lambda$-doubling lines within level $\varv, N_J = 0,16_{33/2}$ observed towards R~Dor.
    The colour map gives the line intensity in mJy~beam$^{-1}$ and the contours show the continuum at levels of 20\%, 40\%, 60\%, and 80\%
   of the peak. We give the observed frequencies not corrected for the source systemic velocity. The full width at half maximum of the
   reconstructing beam is shown by the grey ellipse in the upper left panel.
    }
        \label{fig:mapRDor}
\end{figure*}

The emission region of the OH line within level ${N_J =16_{33/2}}$ observed towards R~Dor is spatially unresolved with a beam
of $0\farcs11 \times 0\farcs16$ and position angle (PA) $\sim -89\degr$. By assuming the brightness distribution to be Gaussian and
fitting it in the image plane using the Common Astronomy Software Applications (CASA) package \citep{McMullin2007},
we find the emission region to have a FWHM of $\sim 70 \pm 20$~mas, and a corresponding
area of ${(6 \pm 3) \times 10^{-3}}$~arcsec$^2$. From a population diagram analysis (Fig.~\ref{fig:popDiag}), we find
a rotational temperature of $2560 \pm 650$~K and a column density of $\sim 2 \times 10^{19}$~cm$^{-2}$,
with an uncertainty of a factor of a few.
The derived values are not as well constrained as for W~Hya because the detected lines span
a smaller range in excitation energy and are not as well fitted by a straight line in the population diagram. Considering
only the two lines observed within 24 hours of each other (within levels $N_J = 14_{29/2}$ and $N_J = 16_{33/2}$),
we find a very high rotational temperature of $23500 \pm 8200$~K.
{The fit obtained using only two transitions is rather unconstrained, however, and strong conclusions cannot be drawn. Nonetheless,
the level populations we infer for the upper levels of these two transitions
might indicate a more complex excitation of OH in R~Dor than our assumption of a single-temperature gas and LTE excitation.}
We also note that R~Dor has been observed to rotate with a velocity of $\upsilon~sin(i) = 1.0$~km/s at the millimetre photosphere \citep{Vlemmings2018}.
Although we do not expect the rotation to affect our analysis of the OH lines significantly, {close examination of the OH channel maps
(Fig.~\ref{fig:mapRDor}) suggests that some effect of rotation is observed despite insufficient resolution.}
{Finally, we did not detect the OH lines within level $v, N_J = 0, 13_{\frac{25}{2}}$ in observations of R~Dor in band~4, with the same set-up as the observations of W~Hya (Table~\ref{tab:lines}).
As shown in the population diagram (Fig.~\ref{fig:popDiag}) this is consistent with the expected line strengths and the sensitivity achieved.}

\subsection{Optical depths and line-flux ratios}

Based on our findings for the rotational temperature, column densities, and emission region, we estimate the optical depths~($\tau$)
of the lines used in the population diagram analysis.
As shown in Table~\ref{tab:ratios}, we find all lines to be optically thin ($\tau \lesssim 0.05$), as assumed in our analysis.

As another diagnosis of the excitation conditions,  we compare the observed ratio between the two lines detected
within a given $N_J$ level to the expected value from optically thin emission and LTE excitation (Table~\ref{tab:ratios}).
The ratios we find for the line pairs observed towards R~Dor are mostly close to the expected values.
The only exception is the pair within level $N_{J} = 12_{23/2}$ at 107.0~GHz,
for which the observed ratio is more than one standard deviation lower than expected.
For W~Hya, two of the four observed pairs (within levels 
$\varv=1, N_{J} = 10_{21/2}$ and $\varv=0, N_{J} = 18_{35/2}$) have ratios that deviate
from the LTE expectation by more than one standard deviation.
This could be a sign of weak maser action in these lines, but the lines are relatively weak and more sensitive observations of a larger
number of lines is required before a robust conclusion can be drawn.

%
%
%
%

\subsection{Transition frequencies}

\begin{figure}[t]
\centering
    \includegraphics[width=8cm]{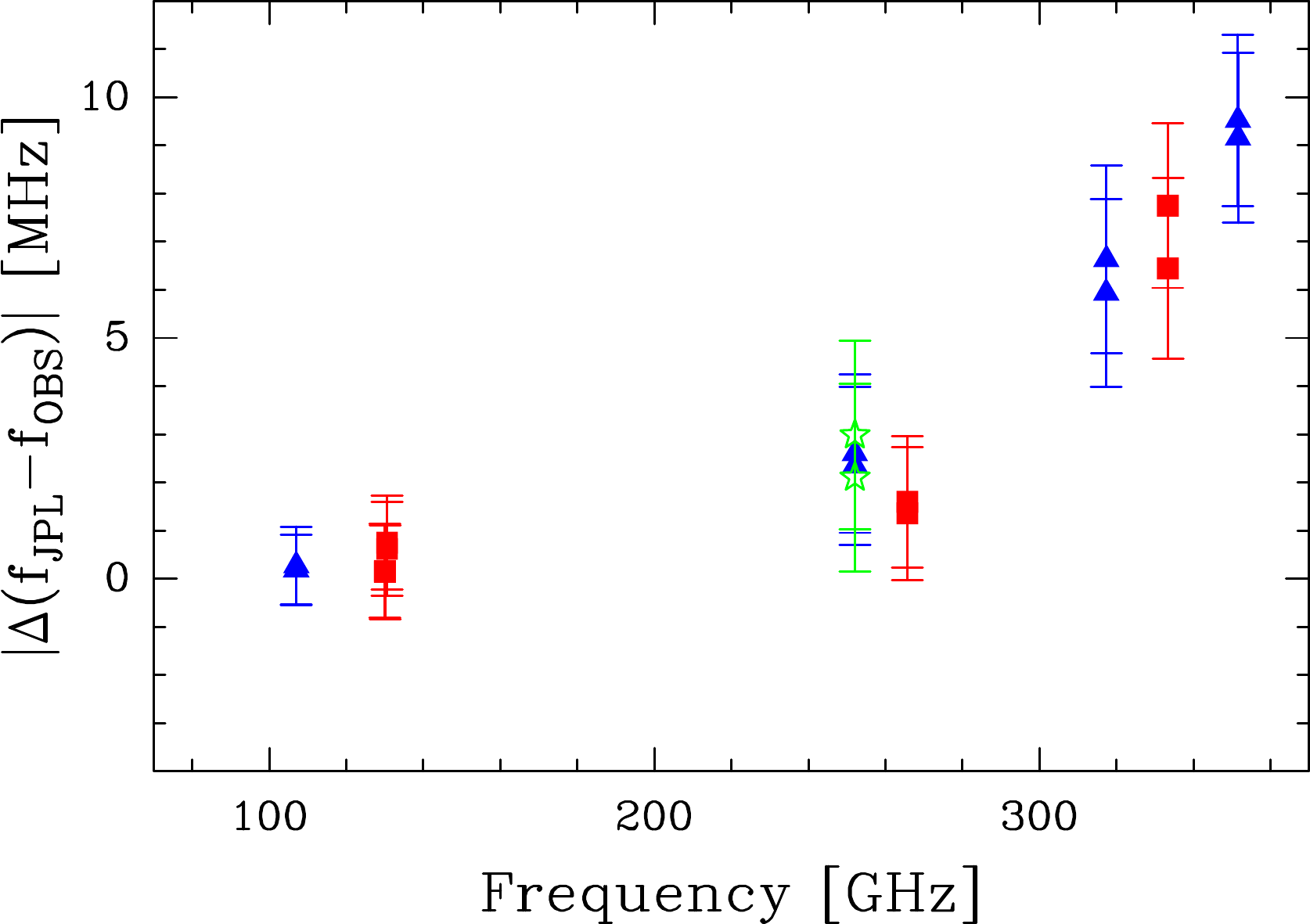}
    \caption{Difference between the predicted frequencies ($\rm f_{JPL}$) and the observed frequencies ($\rm f_{OBS}$) for the observed lines. The blue triangles, red squares,
    and green stars represent the lines observed towards R~Dor, W~Hya, and IK~Tau, respectively.}
        \label{fig:linesFreq}
\end{figure}

\begin{figure}[t]
\centering
    \includegraphics[width=9cm]{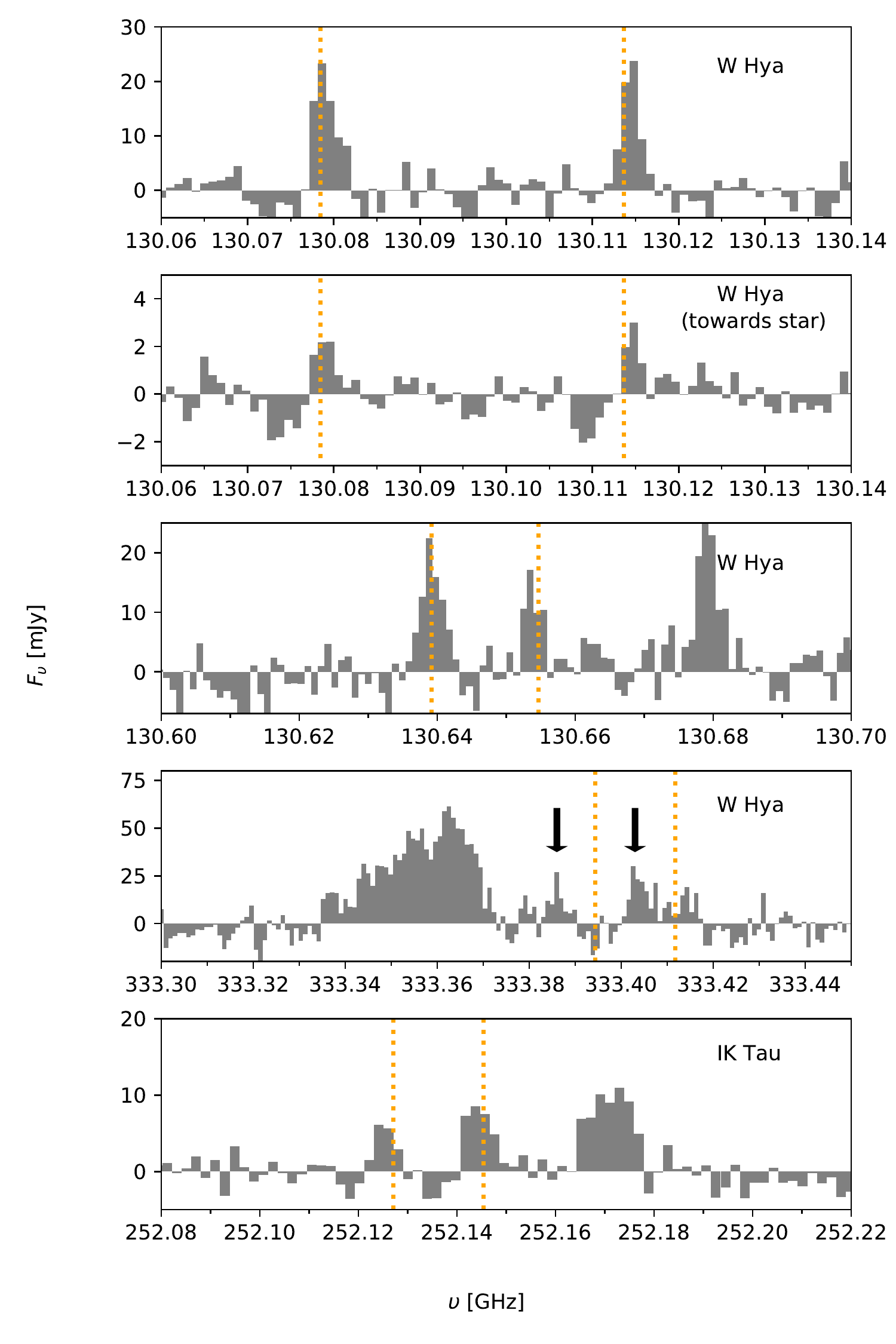}
    \caption{$\Lambda$-doubling lines observed towards W~Hya (used in the population diagram shown in Fig.~\ref{fig:popDiag}) and IK~Tau.
    The spectra are given in rest frequencies using the $\upsilon_{\rm LSR}$ of W~Hya (39.2~km/s)
    and IK~Tau (33.8~km/s).
    The vertical dotted orange lines show the predicted line frequencies from the JPL catalogue.
    The arrows in the second panel from the bottom mark the lines identified as belonging to the OH molecule.
    The lines were extracted from the large region described in the text for W~Hya and from within the central beam of the observations for IK~Tau.
    {Other lines can be seen in the displayed spectra, which we identified  as follows: middle panel, the $J_{K_{\rm a},K_{\rm c}} = 42_{10,32}-43_{9,35}$ SO$_2$ line
    (rest frequency~130.680 GHz); second panel from the bottom, a blend between $J_{K_{\rm a},K_{\rm c}} = 59_{6,54}-58_{7,51}$ and $28_{3,25} - 27_{4,24}$ SO$_2$ lines
    (rest frequencies~333.360 and 333.364~GHz, and possibly also a blend with one line of $^{34}$SO$_2$ and a vibrationally excited line of SO$_2$);
    and bottom panel, $\varv, J_{K_{\rm a},K_{\rm c}} = \varv_1, 7_{4, 3} - 2\varv_2, 8_{5, 4}$ H$_2$O line (tentative identification, rest frequency: 252.172~GHz).}}
        \label{fig:linesWHya_IKTau}
\end{figure}

\begin{figure}[t]
\centering
    \includegraphics[width=9cm]{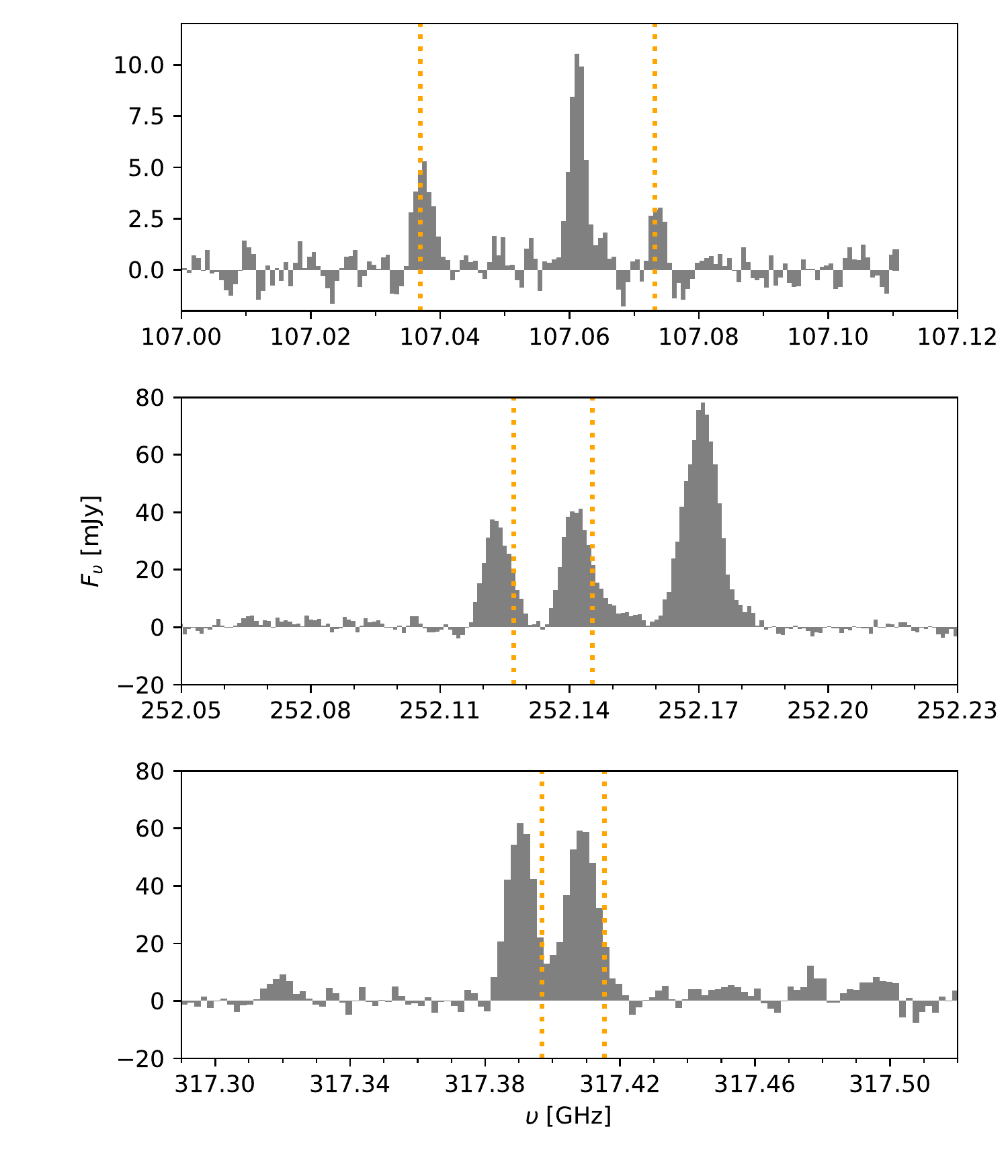}
    \caption{$\Lambda$-doubling lines observed towards R~Dor that were used in the population diagram shown in Fig.~\ref{fig:popDiag}.
    The spectra are given in rest frequencies using the $\upsilon_{\rm LSR}$ of R~Dor
    of 8.5~km/s (see text).
    The vertical dotted orange lines show the predicted line frequencies from the JPL catalogue. The aperture used was twice the size of the band 7 beam.
    {Other lines can be seen in the displayed spectra, which we identified as follows: top panel, the $J_{K_{\rm a},K_{\rm c}} = 27_{3,25}-26_{4,22}$ SO$_2$ line
    (rest frequency: 107.060~GHz); and middle panel, $\varv, J_{K_{\rm a},K_{\rm c}} = \varv_1, 7_{4, 3} - 2\varv_2, 8_{5, 4}$ H$_2$O line (tentative identification, rest frequency: 252.172~GHz).}}
        \label{fig:lines}
\end{figure}

We find that some of the observed OH line frequencies \citep{Drouin2013} differ from the predicted values available in the JPL catalogue\footnote{Available from https://spec.jpl.nasa.gov/} \citep{Pickett1998} by 
more than the uncertainties quoted (Figs.~\ref{fig:linesFreq}, \ref{fig:linesWHya_IKTau}, and \ref{fig:lines}).
This discrepancy does not hamper the line identifications because the separation between the observed lines within a given pair
{is in good agreement with theoretical predictions throughout the OH energy ladder. Further confidence on the line identifications arises from detection of all
 lines expected, given the noise level of the obtained spectra and the derived rotational temperatures.
 Finally, chemical models predict OH to form \citep{Cherchneff2006} and we find no other suitable candidate species
 to explain any of the detected lines when considering typical velocities for gas in these inner regions ($\sim 10$~km~s$^{-1}$).}
In Fig.~\ref{fig:linesFreq} we show that the discrepancy between
observed and theoretical frequencies increases with frequency (and excitation energy).
We derive the observed central frequencies 
by fitting Gaussian functions to the observed lines. We use source velocities with
respect to the local standard of rest, $\upsilon_{\rm LSR}$, equal to 39.2 and 8.5 km/s for W~Hya \citep{Vlemmings2017} and R~Dor (Khouri et al., {\it in prep.}),
respectively. These velocities were derived from modelling the $\varv=1, J=3-2$ CO line produced in the extended atmospheres of these stars.
Based on values reported in literature, we assume an uncertainty of $1.5$~km/s on $\upsilon_{\rm LSR}$ and calculate the uncertainty on the observed
frequency $\sigma = (\sigma_{\upsilon_{\rm LSR}}^2 + \sigma_{\rm fit}^2)^{1/2}$, where $\sigma_{\rm fit}$ is the uncertainty
from the fitting procedure. When $\sigma$ was
smaller than the channel width, it was set to the channel width.

The velocity of the gas in which these lines are produced is not known and it may not be centred at $\upsilon_{\rm LSR}$.
Hence, the derived values of the line centres
may not be representative of the transition frequencies. Moreover, absorption against the stellar continuum may shift the
line centres (Fig.~\ref{fig:linesWHya_IKTau}).
If the lines are produced by gas at the same velocity, the uncertainty on the velocity of the emitting gas
affects each line in a systematic way, and the observed shifts should
be proportional to frequency and different for each source. Since higher excitation lines are probably
produced closer to the star, these might present stronger absorption relative to the lower excitation lines,
and hence show a larger artificial shift. This would produce a trend similar to the one we see.
We note, however, that for W~Hya we see absorption in the transitions within the $\varv, N_J = 0, 13_{25/2}$ level that are redshifted from
the line centre (infalling gas), while the large line-centre shifts observed in the lines within level $\varv, N_J = 1, 17_{35/2}$ are
opposite to those expected if they were caused by a similar absorption {(see Table~\ref{tab:shifts})}.
Given the several sources of uncertainties,
a more precise empirical estimate on the deviation of the line frequencies from predicted values
needs to rely on careful radiative-transfer modelling of the lines. 
These first observations suggest that studies of this type could provide important constraints on
theoretical models of the OH molecule.

 \begin{table}[t]
\footnotesize
\setlength{\tabcolsep}{12pt}
\caption{Observed line-flux ratios compared to predictions for optically thin emission and
excitation following the Boltzmann distribution. The ratios are given by the flux of the higher-frequency line divided
by that of the lower-frequency line (see Table~\ref{tab:lines}).
We also give the estimated optical depth ($\tau$)
for each line when the emission region could be estimated.
The optical depth estimate for the flux from the small region in W~Hya (see text) is given
in parentheses.}              
\label{tab:ratios}      
\centering                                      
\begin{tabular}{@{ }c@{ }c @{\phantom{aa}}c @{\phantom{aa}}c @{\phantom{aa}}c@{\phantom{a}} r@{ }l}
& $\upsilon~{\rm [GHz]}$ & $\varv, N_J$ & Obs. ratio & LTE ratio & $\tau$ \\
\hline
W Hya\\
& 130.1 & $0, 13_{25/2}$ & $0.875 \pm 0.2$ & 0.93 & 0.03 & (0.06)\\
& 130.6 & $1, 10_{21/2}$ & $0.75 \pm 0.26$ & 1.10 & 0.02 & (0.05)\\
& 265.7 & $0, 18_{35/2}$ & $0.71 \pm 0.22$ & 0.95 & \\
& 333.4 & $1, 17_{35/2}$ & $1.2 \pm 0.5$ & 1.06 & 0.002 & (0.01)\\
R Dor & & \\
& 107.0 & $0, 12_{23/2}$ & $0.59 \pm 0.22$ & 0.92 & 0.01 \\
& 252.1 & $0, 14_{29/2}$ & $1.12 \pm 0.07$ & 1.07 & 0.02 \\
& 317.4 & $0, 16_{33/2}$ & $1.04 \pm 0.18$ & 1.06 & 0.02\\
& 351.6 & $0, 17_{35/2}$ & $1.29 \pm 0.31$ & 1.06 & \\
\end{tabular}
\end{table}

\section{Conclusion and outlook}

The detection of highly excited submillimetre lines of the very reactive OH radical in the inner circumstellar environment
of AGB stars provides a new tool that can be used to study the complex dynamics and
chemistry in these inner regions. In particular, a better overview of the excitation structure
of this molecule might help understand whether dissociation in shocks is indeed the main formation pathway.
Moreover, a comprehensive study of the OH mm/sub-mm spectrum towards AGB stars will
provide important constraints on theoretical models of the OH molecule,
and thus to spectroscopic predictions \citep{Drouin2013}. AGB stars are excellent for spectroscopic
characterisation and provide the perfect
laboratories for this type of study \citep[e.g.][]{Cernicharo2018}.

The lines we detected
imply that stronger $\Lambda$-doubling OH lines
should be present in the mm/sub-mm spectrum of O-rich AGB stars. Using the
column density and the rotational temperatures obtained from the population diagram analysis,
we estimate the line pairs within levels $N_J = 8_{17/2}$ (at $\sim 91.2$~GHz), $N_J = 9_{19/2}$ (at $\sim 113.6$~GHz), $N_J = 10_{21/2}$ (at $\sim 137.9$~GHz),
$N_J = 11_{23/2}$ (at $\sim 164.1$~GHz), and $N_J = 12_{25/2}$ (at $\sim 192.0$~GHz) to be respectively $\sim$3.3, 4.6, 5.8, 7.1, and 8.0
times stronger than those within level $N_J = 13_{25/2}$ (at $\sim 130.1$~GHz). Our calculations indicate that these lines should all have $\tau < 1$.

Since OH is a paramagnetic molecule, the magnetic field strength in the line-formation region can
be inferred from the Zeeman splitting experienced by a given line.
This introduces the exciting possibility of measuring the magnetic fields in the atmospheres of close-by
AGB stars using ALMA observations and available techniques.
W~Hya is the best target for  observations of this type
because of the observed line strengths.
The stronger lines that we predict in the $\sim 100$ to $\sim 200$~GHz spectral range are the most suitable
for this type of study.



\begin{acknowledgements}
This work was supported by ERC consolidator grant 614264.
TK, WV, HO, and MM  acknowledge  support  from  the  Swedish  Research  Council.
EDB acknowledges support from the Swedish National Space Agency. This paper
makes  use  of  the  following  ALMA  data:  ADS/JAO.ALMA\#2013.1.00166.S,
ADS/JAO.ALMA\#2016.A.00029.S, ADS/JAO.ALMA\#2016.1.00374.S,
ADS/JAO.ALMA\#2017.1.00075.S, and ADS/JAO.ALMA\#2017.1.00582.S.
ALMA is a partnership of ESO (representing its member states), NSF (USA),
and  NINS  (Japan),  together  with  NRC  (Canada),  NSC  and  ASIAA  (Taiwan),  and  KASI  (Republic  of  Korea),  in  cooperation  with  the  Republic
of  Chile.  The  Joint  ALMA  Observatory  is  operated  by  ESO,  AUI/NRAO,
and NAOJ.
\end{acknowledgements}

\bibliographystyle{aa}
\bibliography{../bibliography_2}

\newpage
\section*{Appendix}

 \begin{table}[t]
\footnotesize
\setlength{\tabcolsep}{12pt}
\caption{Dates of the observations of W~Hya, R~Dor, and IK~Tau in which we detected OH lines
{or obtained a meaningful upper limit (UL)}.
We also give the approximate stellar pulsation
phase when the observations were acquired ($\Phi$). Negative values denote observations obtained at
earlier epochs than the ones we obtained.}              
\label{tab:dates}      
\centering                                      
\begin{tabular}{@{}c@{\phantom{a}}c@{\phantom{a}}c@{\phantom{aa}}c@{\phantom{aa}} c@{\phantom{aaa}} c@{\phantom{}}}
Star & Band & Proj. Code & Lines ($\varv,N_J$) & Date  & $\Phi$\\
\hline
W~Hya & 4 & 2017.1.00075.S & $0,13_{25/2}$; $1,10_{21/2}$ & 17-10-18 & 1.0\\
 & 6 & 2016.1.00374.S & $0,18_{35/2}$ & 17-07-08 & 0.7 \\
 & 7 &  2016.A.00029.S & $1,17_{35/2}$ & 17-11-25 & 1.1\\
R~Dor & 3 & 2017.1.00582.S & $0,12_{23/2}$ & 17-12-18 & 0.7\\
 & 4 & 2016.1.00004.S & $0,13_{25/2}$ (UL) & 17-09-26 & 0.5 \\
 & 6 & 2017.1.00582.S & $0,14_{29/2}$ & 18-01-05 & 0.75 \\
 & 7 & 2017.1.00582.S & $0,16_{33/2}$ & 18-01-06 & 0.75\\
 & 7 & 2015.1.00166.S & $0, 17_{35/2}$ & 15-08-31 & -1.5 \\
IK~Tau & 6 & 2017.1.00582.S & $0,14_{29/2}$ & 18-01-04 & 0.1\\
\end{tabular}
\end{table}

\begin{table*}[t]
\footnotesize
\centering
\setlength{\tabcolsep}{12pt}
\caption{We give the measure differences ($\Delta$(f$_{\rm JPL}$- f$_{\rm OBS}$))
and the corresponding uncertainty ($\sigma_\Delta$)
between predicted (f$_{\rm JPL}$) and observed (f$_{\rm OBS}$) line frequencies. These values
are also displayed in Fig.~\ref{fig:linesFreq}. In Cols. 3 and 4
we indicate the transitions specifying the quantum numbers described
in the text and the positive or negative parity of the energy levels involved. In the last column we give the channel width of
the different spectra.}              
\label{tab:shifts}      
\begin{tabular}{@{}c@{\phantom{aa}} c @{\phantom{aaa}}c @{\phantom{aaa}} c@{\phantom{aa}} c@{\phantom{aa}} c@{\phantom{aa}} c@{}}
Star & f$_{\rm JPL}$ & $\varv, N_{J}$ & $F^{\prime}-F^{\prime\prime}$ & $\Delta$(f$_{\rm JPL}$- f$_{\rm OBS}$) & $\sigma_\Delta$ & Channel width \\
\hline
& [MHz] & & & [MHz] &[MHz] &[MHz] \\
W~Hya \\
& 130078.430 & $0, 13_{\frac{25}{2}}$ & $13^- - 13^+$ & -0.163 & 0.680 & 0.977 \\
& 130113.648 & $0, 13_{\frac{25}{2}}$ & $12^- - 12^+$ & -0.141 & 0.667 & 0.977 \\
& 265734.656 & $0, 18_{\frac{35}{2}}$ & $18^+ - 18^-$ & 1.359 & 1.379 & 0.977 \\
& 265765.312 & $0, 18_{\frac{35}{2}}$ & $17^+ - 17^-$ & 1.600 & 1.367 & 0.977 \\
& 333394.344 & $1, 17_{\frac{35}{2}}$ & $17^- - 17^+$ & 7.751 & 1.705 & 0.977 \\
& 333411.688 & $1, 17_{\frac{35}{2}}$ & $18^- - 18^+$ & 6.448 & 1.873 & 0.977 \\
& 130639.219 & $1, 10_{\frac{21}{2}}$ & $10^+ - 10^-$ & -0.619 & 0.816 & 0.977 \\
& 130654.695 & $1, 10_{\frac{21}{2}}$ & $11^+ - 11^-$ & 0.754 & 0.738 & 0.977 \\
R~Dor \\
& 107036.922 & $0, 12_{\frac{23}{2}}$ & $12^+ - 12^-$ & -0.190 & 0.721 & 0.717 \\
& 107073.172 & $0, 12_{\frac{23}{2}}$ & $11^+ - 11^-$ & -0.273 & 0.812 & 0.717 \\
& 252127.094 & $0, 14_{\frac{29}{2}}$ & $14^+ - 14^-$ & 2.605 & 1.650 & 0.977 \\
& 252145.359 & $0, 14_{\frac{29}{2}}$ & $15^+ - 15^-$ & 2.345 & 1.645 & 0.977 \\
& 317396.750 & $0, 16_{\frac{33}{2}}$ & $16^+ - 16^-$ & 5.941 & 1.854 & 1.953 \\
& 317415.312 & $0, 16_{\frac{33}{2}}$ & $17^+ - 17^-$ & 6.636 & 1.598 & 1.953 \\
& 351593.094 & $0, 17_{\frac{35}{2}}$ & $17^- - 17^+$ & 9.528 & 1.783 & 0.488 \\
& 351611.656 & $0, 17_{\frac{35}{2}}$ & $18^- - 18^+$ & 9.161 & 1.769 & 0.488 \\
IK~Tau \\
& 252127.094 & $0, 14_{\frac{29}{2}}$ & $14^+ - 14^-$ & 2.988 & 1.633 & 1.953 \\
& 252145.359 & $0, 14_{\frac{29}{2}}$ & $15^+ - 15^-$ & 2.109 & 1.660 & 1.953 \\
\end{tabular}
\end{table*}

\end{document}